\documentclass[aps, singlecolumn, showpacs]{revtex4}
\begin{document}
\title{Two-vortex structure of electron, nonlocality and Dirac equation}
\author{S. C. Tiwari \\
Institute of Natural Philosophy \\
1 Kusum Kutir, Mahamanapuri \\
Varanasi 221005, India }
\begin{abstract}
The dimensionless electromagnetic coupling constant $\alpha=e^2 /\hbar c$ may have three interpretations: as the well known ratio between the electron charge radius $e^2/mc^2$ and the Compton wavelength of electron $\lambda_c= \hbar /mc$, as a ratio of two angular momenta since Planck constant has the dimension of angular momentum, and two flux quanta $e$ and $hc/e$. The anomalous part of the electron magnetic moment together with the unified picture of the three interpretations of $\alpha$ is suggested to have deep physical significance. The electric charge is proposed to be a new quantum of flux that leads to a new model of the electron envisaging a two-vortex structure. In analogy with quantum conditions we postulate sub-quantum conditions applicable in a region of the order of $\lambda_c$ replacing $\hbar$ by a universal constant $f=e^2 /2\pi c$ and apply it to Dirac equation in internal space that gives rise to the anomalous magnetic moment of electron. Dirac spinor and 2-spinor representation for vortex structure of electron in the single particle Dirac framework are discussed. The role of sub-quantum rules and the internal variables for developing the present ideas is also debated. A critical discussion on the past attempts to give fundamental importance to magnetism and flux quantum is given to delineate the new ideas in the present work.

\end{abstract}
\maketitle

\section{\bf{ INTRODUCTION}}

Relativistic theory of spinning electron was independently given by Frenkel \cite{1} and Thomas \cite{2} prior to the most celebrated Dirac's relativistic equation \cite{3}.
In Section 5 of his paper Thomas considered Abraham's model of electron assumed to be a spinning sphere of electric charge distribution, and concluded that for a meaningful notion 
of the structure of electron, theory of general relativity would be necessary. Point charge in Dirac's theory and classically non-describable spin and magnetic moment that emerge from it have been debated since the work of Dirac \cite{3}. The role of a characteristic length scale of the order of Compton wavelength $\lambda_c =\hbar /mc$ has been recognized in the physical interpretation of Dirac's single particle theory. Ironically the classical electron charge radius $r_e =e^2 /mc^2$ though much smaller than $\lambda_c$ does not seem to be important in quantum theory. Finite magnetic moment and spin as some kind of rotation entail finite spatial extension for electron, however all efforts to model electron in this spirit have failed. Recent advances in quantum electrodynamics (QED) and the Standard Model (SM) of particle physics have proved their empirical success to an extremely high precision. Physics beyond the SM is primarily inspired by the foundational problems: elusive Higgs boson and Higgs mechanism, large number of adjustable parameters, the origin of electric charge quantization, and the nature of infinities and their handling via renormalization procedure. While sub-structure of nucleons and hadrons constitutes an important ingredient of the SM, the leptons are assumed elementary. Even in some of the speculations beyond the Standard Model the lepton structure envisaged is not in the spirit of the original electron models. For example, leptoquark signature indicated in HERA experiments in 1997 critically reviewed in \cite{4} has not been confirmed since then. We have argued \cite{4} that insurmountable difficulties in the extended electron models and the foundational problems in QED arise due to the lack of the understanding on the nature of the electric charge.

Recent new insight that electric charge could be interpreted as a quantum of flux akin to the flux quantum $hc/e$ leads to a new model of the electron: this model is developed in a two-vortex approximation neglecting $\alpha^2$ and higher order terms in the expression of the magnetic moment. A new framework for the internal dynamics of electron is suggested based on what we term as sub-quantum conditions. Proposed sub-quantum theory is based on the realization of a remarkable fact that QED correction to Bohr magneton does not depend on Planck constant that is believed to be a typical signature of quantum world. Obviously QED effect cannot be classical and Planck constant does not appear in it, therefore it is logical to seek a sub-quantum origin. A pair of spinors to represent the two vortices is discussed: an almost point vortex rotates around a core vortex in this model. Obviously the mathematical description would not be easy; a recent work on orientable objects by Gitman and Shelepin \cite{5} could prove useful to develop the formalism.

To set the motivation for extended electron model the problem of nonlocality and the electron magnetic moment is discussed in the next section following the well known arguments given by Dirac \cite{3}. Heuristic arguments leading to the two-vortex structure of electron are presented in Sec.III. Sub-quantum conditions are formulated in Sec.IV, and their implication to account for the anomalous magnetic moment of electron is also discussed. A simple straightforward generalization of Dirac equation to 8-component spinor representation of electron is debated, and a  more radical possibility in terms of 2-spinor equations corresponding to two vortices is also discussed. In the last section the new elements in the present electron theory are delineated and placed in the context of past apparently failed attempts \cite{6, 7, 8, 9} that gave fundamental role to magnetism and magnetic flux quantum towards the unification goal.

\section{\bf{Nonlocality and electron magnetic moment}}

Let us begin with Dirac Hamiltonian for free electron
\begin{equation}
 H=c {\bf{\alpha .p}} +\beta mc^2
\end{equation}
where the $4\times 4$ matrices $\bf{ \alpha}$ and $\beta$ in one of the usual representations could be expressed in terms of $2\times 2$ Pauli spin matrices $\bf{\sigma}$ and the unit matrix I
\begin{equation}
 \bf{\alpha} =\left(\begin{array}{cc} 0 & \bf{\sigma} \\ \bf{\sigma} & 0\\
                     
                    \end{array}\right)
\end{equation}
\begin{equation}
 \beta =\left(\begin{array}{cc} I & 0 \\ 0 & -I\\
                 
                \end{array}\right)
\end{equation}
In Section 69 of \cite{3} the time derivative of position operator $\bf {r}$ is derived using the Heisenberg equation of motion
\begin{equation}
{ \bf{\dot{r}}}=c\bf{\alpha}
\end{equation}
Since $\bf{\alpha}$ has eigenvalues $\pm 1$, the velocity eigenvalues are $\pm c$. This result contradicts the observed motion of electron; Dirac offers a physical explanation invoking the Heisenberg uncertainty principle. To measure velocity of an electron one has to determine exactly its position at two infinitesimally close time points thereby making the momentum indeterminate. Thus according to Dirac, 'This means that almost all values of the momentum are equally probable, so that the momentum is almost certain to be infinite. An infinite value for a component of mementum corresponds to the value $\pm c$ for the corresponding component of velocity.'

To gain further insight the time variation of the velocity of electron is calculated from the operator equation, for example, for the x-component we have
\begin{equation}
i\hbar \dot{\alpha_1} =\alpha_1 H- H\alpha_1
\end{equation}
Solving for $\alpha_1$ and integrating again we have
\begin{equation}
 c\alpha_1 =v_x +v^{o} _x
\end{equation}
\begin{equation}
 x=v_x t +a_1 +x^o
\end{equation}
\begin{equation}
 v_x = c^2 p_1 H^{-1}
\end{equation}
\begin{equation}
 v^{o} _x =\frac {1}{2} ic\hbar \dot{\alpha_1} (0)e^{-2iHt/\hbar} H^{-1}
\end{equation}
Here $\alpha_1 (0)$ and $a_1$ are integration constants. The constant part $v_x$ in the expression (6) corresponds to the observed velocity of electron as shown by Eq.(8). The oscillatory part $v_{x} ^o$ given by expression (9) has instantaneous value equal to the velocity of light, however for the time intervals large compared with the time period of its oscillation of the order of $2mc^{2} /h$ the average velocity would be equal to $v_x$. It is also important to note that rather than a particle localized at a point the position of electron possesses an oscillatory part $x^o$ in Eq.(7) with a spread of the order of Compton wavelength
\begin{equation}
 x^o = \frac {1}{2} i\hbar v^{o} _x H^{-1}
\end{equation}

One can see that since $\alpha$ matrices do not commute, the velocity components of electron cannot be measured in all directions. Dirac does not discuss this, however Fock \cite{10} considers the lack of physical interpretation of the velocity operator as a defect in Dirac's theory. In a significant advancement Foldy and Wouthuysen \cite{11} showed that performing a canonical transformation on the free particle Dirac Hamiltonian (1) a new representation (F-W representation) led to the separation of positive and negative energy states in terms of 2-spinors. Authors derive a new position operator and calculate its time derivative to obtain the physical interpretation arriving at the conventional velocity operator \cite{11}. The new position operator is termed as mean-position operator and as explained by them in footnote 7 of their paper such an operator had earlier been independently discussed by others, e. g. Pryce in 1935 and Newton and Wigner in 1949. Recently a useful extension to the F-W transformation for relativistic particle interacting with external fields has been made by Silenko \cite{12}. Gill, Zachary and Alfred \cite{13} make an important point regarding nonlocality: there is implicit time nonlocality in Dirac equation. The velocity eigenvalues equal to $\pm c$ and zitterbewegung are interpreted as oscillations between the past and the future at the speed $c$ of a spatially extended particle forced to appear as a point particle in Dirac theory. A careful reading of Dirac's arguments \cite{3} and F-W interpretation in \cite{11} would show that reluctance to ascribe spatial extension to the electron results into the implicit time nonlocality in the form of the oscillatory time period of the order of $2mc^2 /h$ supporting the conclusion arrived at in \cite{13}.

Localization of electron wavepacket in a spatial region of less than the Compton wavelength necessarily results into the interference between negative and positive energy states and the one-particle picture of Dirac theory is inadequate; second quantization or QED in a way resolves this issue. Interestingly physical interpretation of bare charge and renormalized charge due to vacuum polarization in QED invokes \cite{14} electron and positron cloud within a distance of $\lambda _c$.

In the F-W representation \cite{11} one defines a mean position operator and a mean spin operator such that the mean spin is a constant of motion unlike the Dirac representation in which spin is not a constant of motion. Due to the finite extension of the electron around the mean position of the order of $\lambda_c$ interaction with external electromagnetic field is expressed in terms of a multipole expansion in \cite{11}; the dipole term gives rise to the magnetic moment. Note that the role of the oscillatory position (10) is acknowledged by Dirac in Section 70 of \cite{3} while discussing spin. Huang making a departure from ambiguous role accorded to nonlocality examines the role of the length $\lambda_c$ visualizing electron spin and magnetic moment arising from a circular motion in a radius equal to the Compton wavelength \cite{15}. Thus zitterbewegung is assumed to be a physical internal motion of electron. Relatively recently van Dyck et al \cite{16} reporting precision measurement of the magnetic moment of free electron ponder over a kind of quasi-orbital radius of the order of $\lambda_c$ for a point charge undergoing circular zitterbewegung.

Experimental study on the hyperfine splitting of energy levels in Zeeman effect in hydrogen and deuterium carried out by Nafe, Nelson and Rabi \cite{17} showed significant discrepancy between the measured and the calculated hyperfine structure separation frequencies. It is now understood in terms of anomalous magnetic moment of electron deviating from the exact Bohr magneton predicted by Dirac theory
\begin{equation}
 \mu _B =\frac{e\hbar}{2mc}
\end{equation}
The QED calculated value of the electron magnetic moment in the powers of the fine structure constant $\alpha =e^2 /\hbar c$ neglecting higher order terms is  given by
\begin{equation}
 \mu _e =\mu _B [1+\frac{\alpha}{2\pi} -\frac{0.328478444 \alpha ^2}{\pi}]
\end{equation}
Precision experimental tests have been found to agree extremely well with QED calculations.

A natural question is whether anomalous magnetic moment of electron can be obtained in Dirac picture of single particle. Pauli in 1941 introduced additional terms in Dirac equation \cite{18} which could give arbitrary magnetic moment to the Dirac particle, of course, remarking that in the case of electron such a term was not required as electron was known to have magnetic moment exactly equal to Bohr magneton $\mu_B$. Lateron Foldy investigated the most general additional interaction terms \cite{19} allowed in the Dirac equation imposing the conditions of Lorentz covariance and gauge invariance assuming the interaction to be linear in the electromagnetic potentials. Let us briefly discuss Dirac equation for interacting electron. For free particle given the Hamiltonian (1) Dirac equation reads
\begin{equation}
 i\hbar \frac{\partial \psi}{\partial t} =(-i\hbar c {\bf{\alpha.\nabla}} +\beta mc^2 )\psi
\end{equation}
where $\psi$ is a 4-component column vector, i. e. Dirac spinor. Manifestly covariant form of Dirac equation is obtained employing Dirac gamma matrices $\gamma^\mu,(\mu =0, 1, 2, 3)$
\begin{equation}
 (i\hbar \gamma^\mu \partial_\mu -mc)\psi =0
\end{equation}
The notation used is $p^\mu =i\hbar \partial ^\mu, \partial^\mu =(\frac{\partial}{\partial x^0}, -{\bf{\nabla}})$ and $x^\mu =(x^0=ct, {\bf{x}})$. Interaction with the electromagnetic field in analogy with classical electrodynamics is described replacing $p^\mu$ by $p^\mu -\frac{e}{c} A^\mu$. where $A^\mu$ is electromagnetic potential 4-vector. Eq.(14) is transformed to
\begin{equation}
 \gamma^\mu (i\hbar \partial_\mu -\frac{e}{c} A_\mu)\psi =mc\psi
\end{equation}
Pauli added a term proportional to $\gamma^\mu \gamma^\nu F_{\mu\nu}$ that would give rise to an anomalous magnetic moment. Here the electromagnetic field tensor has the usual definition
\begin{equation}
 F_{\mu\nu} =\partial_\mu A_\nu -\partial_\nu A_\mu
\end{equation}
It is evident from Eq.(15) that it represents a point charge interacting with the field, however it is well known that when worked out in detail additional interaction terms arise implicitly present in the form (15). In the nonrelativistic approximation the dominant terms are the Pauli magnetic moment interaction and the spin-orbit coupling. In the F-W representation the physical interpretation becomes more transparent \cite{19}. Foldy's generalized equation given in \cite{19} contains the usual point charge interaction, however the magnetic moment gets modified by an anomalous part and the Darwin term comprises of three parts: a Darwin term similar to that found in Dirac equation (15), a contribution due to the anomalous magnetic moment, and an intrinsic Darwin term. It is possible to assume the arbitrary expansion coefficients in Foldy's generalized Dirac equation to get the electron magnetic moment that agrees with the expression (12), however as pointed out by Foldy the interpretation of intrinsic Darwin term is problematic in the light of the infrared divergencies in QED.

To conclude this section: negative energy states, nonlocality and infinities characterize the difficulties associted with Dirac theory. Though renormalization method succeeds in getting sensible physics out of the divergencies in QED, and there does not seem to be any urgency or crisis in theoretical physics for the revival of the models based on extended objects for elementary particles, it is true that the progress in quantum field theory has tended to become stagnant. In the present paper the nonlocal aspects associated with two lengths $\lambda_c$ and $r_e$ are investigated based on the single-particle Dirac equation in a new framework as explained in the next section: the idea is that a radical departure at fundamental level in electron theory might ultimately explain the deeper hidden physics behind the success of the renormalization method in QED.

\section{\bf{ STRUCTURE OF THE ELECTRON}}

Extended electron models beginning with Thomson's model faced the insurmountable difficulty: electric charge $e$ assumed to be elementary when viewed as a continuous distribution over a spatial region (surface or volume) cannot be stable as  distributed like charges would repel each other. Postulating cohesive Poincare stress of unknown origin or introducing some kind of non-electromagnetic force to stabilize the electron have been of limited success since the original idea to achieve unification and simplicity gets lost. In an exhaustive discussion \cite{4} we have reviewed this problem and argued that a clue to make progress in the electron model would come from understanding the nature of electric charge itself: What is charge? Does it have a purely mechanical origin? We know that besides the electric charge electron has spin angular momentum $\frac{\hbar}{2}$, a magetic moment $\mu_e$, and indirectly two length scales characterize it, namely the Compton wavelength and the classical electron charge radius. In the preceding section it has been pointed out that the length $\lambda_c$ is implicitly related with spin. It is logical to imagine electron charge radius also to be related with some sort of spinning motion. Let us focus on the first two terms in the expression (12) for $\mu_e$ and re-write it in the following form
\begin{equation}
 \mu_e =\frac{e}{mc} [\frac{\hbar}{2} +\frac{f}{2}]
\end{equation}
\begin{equation}
 f= \frac{e^2}{2\pi c}
\end{equation}
The second term in the square bracket of (17) is suggestive of interpretating it as angular momentum, and it being a fraction of spin $\frac{\hbar}{2}$ we termed it as fractional spin of electron in \cite{4}. Remarkably $\frac{f}{2}$ does not contain Planck constant and depends solely on electric charge and vacuum velocity of light, therefore the hypothesis was put forward that electric charge was a manifestation of fractional spin \cite{4}. Another important hint for this interpretation arises examining the structure of the fine structure constant $\alpha$: it is a dimensionless constant and it is well known that it could be viewed as a ratio of electron charge radius $e^2 /mc^2$ and Compton wavelength $\hbar /mc$. Curiously one could also interpret $\alpha$ as a ratio of two angular momenta $e^2 /c$ and $\hbar$. Is there a connection between this fact and the aforementioned hypothesis? Linking electric charge with (fractional!) spin amounts to a purely mechanical interpretation of the electromagnetic phenomena. This idea goes beyond the equivalence sought between the electromagnetic energy and the mechanical energy or rest energy in all electron models considered till date. The sign of electricity (positive or negative) could be easily related with the sense of rotation that gives rise to angular mementum $f/2$, however $f$ contains $e^2$ not $e$ that makes the physical interpretation less transparent. 

A recent breakthrough \cite{20} has occurred with a possible third interpretation of the fine structure constant. Recalling that magnetic flux quantum is $hc/e$, it immediately follows that $\alpha$ could be viewed as a ratio of two flux quanta such that electric charge is a new quantum of flux. A natural abstraction of the magnetic flux quanta drawing an analogy with vortices in rotating fluid would be that these correspond to the strengths of the quantized vortices. Once electric charge is assumed to be a quantized vortex, the problem of the quantization of charge postulating magnetic monopole becomes superfluous. Note that the spinning vortex is assumed to exist in the abstract spacetime fluid: it should not be incomprehensible since spin has to do with something rotating and without introducing artificial constructs it is simplest to assume that something to be spacetime.

A unifying picture of three interpretations hidden in the fine structure constant, namely the ratio of two fundamental lengths, the ratio between two spin angular mementa, and the ratio between two flux quanta is achieved in a proposed two-vortex structure of the electron. A central vortex C and an orbiting vortex O having the core radii respectively of the order of $\lambda_c$ and $r_e$ and circulation strengths of $\Gamma _g$ and $\Gamma _e$ constitute the electron (positron). Intrinsic angular mementum or spin of the structure is given by the term in square bracket in (17) and dividing by $\hbar$ we can define the vortex strengths as dimensionless numbers
\begin{equation}
 \Gamma _g =\frac{1}{2}
\end{equation}
\begin{equation}
 \Gamma _e =\frac{\alpha}{4\pi}
\end{equation}
The sign of circulation determines the sign of charge such that vortex-vortex repel and vortex-antivortex attract. To the outside observer the rotating vortex O appears as a source of point charge $e$. The electrostatic Coulomb field is an effect of time varying flux quantum $e$ qualitatively similar to Jehle's argument \cite{8}. Electron has two distinct internal configurations $O_-C_+$ and $O_-C_-$, here suffix $-$ or $+$ denotes vortex or antivortex. Positron is represented by $O_+C_+$ and $O_+C_-$ in this scheme. Since the vortex C accounts for the spin of the electron, 2-vortex structure in which the vortex O rotates around C admits the intriguing possibility of spin and charge separation as O could become free from the circular orbit; it also implies that electron with spin up and spin down are distinct states due to different internal structures \cite{20}.

At least three questions have to be addressed before this model could be developed further. Firstly one may object to giving fundamental significance to the individual terms in the QED calculated magnetic moment: $\mu_B$ and $\mu_B \alpha/2\pi$. Let us go back to the Dirac equation that in a neat form predicts magnetic moment of electron equal to Bohr magneton, and the spin of $\hbar /2$ is related exactly with it. The discrepancy in the hyperfine structure between that expected from theory and observed experimentally in \cite{17} was tentatively explained by Breit invoking anomalous magnetic moment of electron \cite{21}; subsequent developments in QED led to precise calculations of higher order terms in the power of the fine structure constant. However an interesting remark made by Breit is worth quoting: 'It is not claimed that the electron has an intrinsic magnetic moment. Aesthetic objections could be raised against such a view. The only object of this note is to point out that the evidence considered above does not disprove a small $\mu_e$ of the order $\alpha \mu_0$'. Here $\mu_0$ is Bohr magneton and $\mu_e$ anomalous part in the electron magnetic moment in Breit's notation. Since the first term in (12) embodies the beautiful relationship between spin and magnetic moment a la Dirac spinor theory, it would be aesthetically pleasing to seek similar relationship pertaining to anomalous part in the magnetic moment of electron.

The second question relates with the longstanding problem of the rest mass (energy) $mc^2$ of the electron. It is ironical that though the elementary particles or fields could be endowed with non-zero angular momentum or spin the expected corresponding rotational energy is absent in the dynamics. It was argued recently \cite{22} that half of the photon  energy $h\nu$ be identified with the spin energy of the photon. In an analogous manner for the electron the rest energy $mc^2$ is suggested to originate from the internal rotation. A plausible postulate is to  ascribe half of the rest energy each to the spin $\hbar /2$ and $f/2$ \cite{20}. In a simplified rotating disk model for two vortices it is found that the length scales of $\lambda_c$ and $r_e$ appear respectively for them. There is an interesting discussion on spin energy for elementary particles having internal structure by Finkelstein \cite{23}.

Finally the issue of fractional spin. What does this mean for free electron? Primarily the fractional spin in our model represents the spin $f/2$ that we propose for electron corresponding to the second term in expression (12) for the electron magnetic moment; fractional signifies merely the fact that its magnitude is a fraction of $\hbar$. Since the model envisages a circular spatial region in a plane such that the propagation of electron is normal to it, the idea of anyon having fractional spin appears  quite attractive; in fact, in the monograph \cite{4} we presented this interpretation as one of the most likely ones. The importance of the Euclidean group E(2) in this connection has been lucidly described in a comprehensive article by Kastrup \cite{24}; though his emphasis is on the fractional orbital angular momentum in quantum optics a great deal of discussion and the references cited in it have wider ramifications. Taking the group E(2) for electron model one could seek a group theoretical quantization, however the recent idea that electron consists of two vortices \cite{20} as explained above indicates that rather than E(2) one should look for $E(2)\times E(2)$ for this problem. Assuming each vortex as an independent object, would it be possible to seek spinor representation for each of them? Numerical value of $f$ as a fraction of $\hbar$ does not necessarily imply a fractional spin object and anyon statistics in this case. In the next section the standard quantum conditions are considered afresh to develop a new approach for internal dynamics of electron.

\section{\bf{ Sub-quantum conditions and Dirac equation}}

A point particle could be assigned internal attributes postulating hypothetical internal spaces, for example, spin or isospin space. A more concrete picture in which extended spatial structure represents internal properties becomes difficult in view of the relativistic and the quantization requirements. Yukawa developed a nonlocal quantum field theory in which point particle is replaced by a finite radius object \cite{25}. In a  beautiful classification scheme depending on the symmetry groups Finkelstein \cite{23}  considered rigid internal structure of particles and found that spin-half Dirac spinors are allowed. Recent article \cite{5} highlights the unsatisfactory progress in dealing with extended orientable system, and suggests a relativistic framework: in particular, it throws light on the four spinors and argues that of the four types two are present in the explicit form in Dirac equation, and the remaining two are given by complex conjugation and the sign of the mass term. Actually the relativistic equation in \cite{5} is 8-component spinor equation and splits into two Dirac equations with opposite sign of the mass term. Note that in the nonlocal Yukawa theory the spinor field is also an 8-component spinor. A short comment at the end of Yukawa's paper shows that 8 components are not imperative, and Fierz \cite{26} doubts whether the nonlocal field of Yukawa has indeed a finite radius of the particle. In the context of time reversal symmetry Biedenharn \cite{27} also proposed an 8-component spinor equation for Dirac electron; see \cite{28} for further elucidation.

The above cited literature shows that retaining the basics of Dirac theory generalizations are possible for extended particle or internal degrees of freedom. For a planar system E(2) symmetry group admits fractional quantization for phase space \cite{24}; here E(2) is a direct product group of rotations SO(2) and translations T(2). Description of an orientable rigid body in 2-dimensional space requires $E(2) \times E(2)$ as illustrated in \cite{5}. Obviously even for a single vortex that has finite area the symmetry group is $E(2) \times E(2)$, therefore, 2-vortex electron model would be a complicated problem having further requirements. The crucial point is the recognition that QED correction to Bohr magneton is a term that does not depend on Planck constant, namely the second term in the expression for the electron magnetic moment (12) equal to $ef/2mc$. It is generally accepted that classical and quantum effects are differentiated by the absence or the presence of $\hbar$. Is QED effect classical? Considering the electron model in which we are envisaging the internal structure of electron, the dynamics refers to sub-electron and cannot be classical. It is logical to suggest that it could be obeying new set of what we call sub-quantum conditions in which $f$ replaces the role of $\hbar$. Recall that quantum conditions can be introduced in analogy with classical dynamics using the Poisson bracket, e. g. Chapter 4 in \cite{3}. One can postulate sub-quantum conditions in analogy with quantum conditions replacing $\hbar$ by $f$ for the internal space. The quantum Poisson bracket of any two variables $u$ and $v$ is defined to be
\begin{equation}
 uv -vu =i\hbar [u, v]
\end{equation}
It is instructive to quote Dirac following this definition on p. 87 in \cite{3}, '... in which $\hbar$ is a new universal constant. It has the dimension of action. In order that the theory may agree with experiment, we must take $\hbar$ equal to $h/2\pi$, where $h$ is the universal constant that was introduced by Planck, known as Planck's constant'. Note that the universal constant in the quantum condition (21) is identified with the Planck constant divided by $2\pi$ for dimensional reason and agreement with experiments. Since $f$ has the dimension of action, and comprises of universal constants $e$ and $c$, we postulate sub-quantum condition
\begin{equation}
 uv - vu =i f [u, v]
\end{equation}
 For the canonical coordinates and momenta we get the fundamental sub-quantum conditions
\begin{equation}
 q p - p q = if
\end{equation}
Following the standard quantum mechanical approach it is straightforward to write down the 4 - momentum operator $p^q =if \partial ^a$ in the internal spacetime for sub-quantum theory. To distinguish internal variables we use the index $a =0, 1, 2, 3$ instead of $\mu$. Dirac equation (14) for free particle becomes
\begin{equation}
 (if \gamma^a \partial _a -mc) \chi (x^a, x^\mu) = 0
\end{equation}
Four- spinor $\chi$ is a function of both spacetime variables, however let us assume that it depends on $x^a$ for the present moment. To describe interaction with the electromagnetic field we assume
\begin{equation}
 p^a \rightarrow p^a - \frac{e}{c} A^a
\end{equation}
This assumption rests on a weak analogy, and will have to be modified in a more complete theory as we discuss below, however once we assume (25) an equation analogous to (15) is obtained
\begin{equation}
 \gamma^a (if \partial_a - \frac{e}{c} A_a ) \chi =mc \chi
\end{equation}
Perforimg the standard calculations one can obtain a magnetic moment term equal to $ef/2mc$ and a length scale of $r_e /2\pi$ instead of $e\hbar /2mc$ and $\lambda_c$ in Dirac theory respectively. More important is the spin half representation embodied in the 4-spinor $\chi$, in spite of spin angular momentum being $f/2$. 

Thus composite structure of electron is represented by two Dirac spinors $\psi$ and $\chi$ with a naive description given by equations (15) and (26), and corresponds to a kind of factorized two particle model in which $\psi$ and $\chi$ depend on external and internal spacetime variables respectively. In view of this simplification it is not difficult to show the consistency of the present formulation accounting for the interaction term having total magnetic moment (17). Adopting the usual approach transforming first-order Dirac equation to a second-order form Eq.(26) is transformed to
\begin{equation}
\gamma^a (if \partial_a - \frac{e}{c} A_a ) \gamma^b (if \partial_b - \frac{e}{c} A_b ) \chi =m^2c^2 ~ \chi
\end{equation}
To simplify it we make use of the sub-quantum conditions (23) and derive the relation
\begin{equation}
 [p_a, ~ A_a] = if \frac{\partial A_a}{\partial x^a}
\end{equation}
The properties of Dirac gamma matrices and (28) finally yield
\begin{equation}
 (p^a - \frac{e}{c} A^a)(p_a -\frac{e}{c} A_a) \chi + \frac{1}{2} \frac{ef}{c} \sigma^{ab} F_{ab} \chi = m^2 c^2 \chi
\end{equation}
The second term $\frac{1}{2} \frac{ef}{c} \sigma^{ab} F_{ab}$ is easily recognized to be the anomalous magnetic moment interaction term, and from $\psi$ Dirac equation we get Bohr magneton term thus establishing consistency with the known facts.

Note that equations for $\psi$ and $\chi$ can be superposed introducing $8 \times 8$ matrices consisting of $\gamma^\mu$ and $\gamma^a$ and
8-component spinor comprising of $\psi$ and $\chi$ keeping track of quantum and sub-quantum conditions. However we ask the following question.
Is there a better mathematical framework for incorporating internal space and sub-quantum conditions in Dirac formalism? In the literature that deals with internal degrees of freedom, let us first consider Yukawa's nonlocal theory \cite{25}: two sets of independent variables are defined using the spacetime coordinates $x^\mu$, and assumed plane wave representation in an inertial reference frame in which the particle is at rest leads to a physical interpretation such that the internal coordinates determine the size of the particle. On the other hand Finkelstein \cite{23} keeps interal coordinates $q_a (a=1, .... N)$ independent of the spacetime coordinates $x^\mu$; however under Lorentz transformations the internal coordinates are assumed to change in a specified manner, in particular, a spacetime translation leaves $q_a$ invariant. We adopt the following view: there is a universal spacetime continuum, and the classification of internal variables or internal degrees of freedom depends on the applicable constraints or dynamical conditions, namely, classical, quantum or sub-quantum. A classic example in which internal space is redundant in dynamics is that of a rigid body: the internal potential remains constant and the rigid body dynamics in 3-dimensional space can be described by six independent generalized coordinates. One may define two reference frames, space-fixed and body-fixed, and describe the system by two sets of coordinates with appropriate transformation laws for body-fixed coordinates with respect to the space-fixed reference frame \cite{5}. An interesting problem is that of two body dynamics in a mutual central force: it can be reduced to a one-body problem in which center of mass motion can be ignored and the difference position vector of two bodies only enters the dynamics. Hydrogen atom and Kepler's planetary problems are similar, however they differ due to the quantum conditions imposed on the relative coordinates treated as internal variables in the former. Though the center of mass motion of an atom usually is of no consequence quantum mechanically, in an interesting interplay between external and internal motion it could have important bearing on some phenomena, for example, Doppler effect \cite{29}. Fermi in \cite{29} argues that the assumption that the phase of the electromagnetic plane standing wave remains constant as the portion of space where electron moves is very small cannot remain valid, and its dependence on the center of mass coordinates has to be taken into account. Naive looking Eq. (24) embodies this deeper foundation: the subquantum conditions (23) define internal space that essentially lies within the region of the order of Compton wavelength.

Obviously the separation into $x^\mu$ and $x^a$ is not rigid, and treating the spinors $\psi$ and $\chi$ independently to describe electron has to be an approximation to a more complete theory. Note that in the light of the current status of the constancy of the universal constants, i. e. no definite evidence for their variability, it is reasonable to assume that $\hbar$ and $f$ are universal constants defining respectively the quantum and sub-quantum dynamics. The mass or rest energy or the intrinsic energy of electron could depend on the internal structure of electron, in fact, as already discussed its physical interpretation in terms of the electromagnetic self-energy has been investigated by numerous physicists and  has not been successful. Here a radically new perspective is possible: assume it to be zero and see if it could arise from internal motion. If it succeeds it may offer a different approach to the issue of negative energy states in Dirac theory. Next question is that of electric charge: if $f$ is universal constant the charge parameter entering in its expression has to be a constant, however it is not necessary that the same parameter should appear in Eq. (25). We may introduce a charge parameter $e_s$ for coupling with the external electromagnetic field. In the present work, the quantum world is a kind of classical approximation to the sub-quantum world: the ratio $f/\hbar$ is very small equal to $\alpha /2\pi$, and we may write for the two lengths
\begin{equation}
 \frac{\hbar}{mc} = \frac{e^2}{mc^2} + \sum ~  n\frac{e^2}{mc^2}
\end{equation}

Here summation is from $n = 1, ... 16$. Regarding the mass term we first illustrate the idea by a simpler example of scalar field $\phi$ that depends on two sets of variables $x^\mu$ and $x^a$. Now using $p^\mu p_\mu +p^a p_a$ we get the second order wave equation
\begin{equation}
 (\hbar^2 \partial^\mu \partial_\mu + f^2 \partial^a \partial_a ) \phi = 0
\end{equation}
Under the assumption of the separation of variables $x^\mu$ and $x^a$, the function $\phi$ can be written as a product $\phi(x^\mu) \phi(x^a)$ and the equation for, let us call it physical field $\phi(x^\mu)$ becomes
\begin{equation}
(\hbar^2 \partial^\mu \partial_\mu + K^2 ) \phi(x^\mu) = 0
\end{equation}
setting
\begin{equation}
 f^2 \partial^a \partial_a \phi(x^a) = K^2 \phi(x^a)
\end{equation}
Assuming the separation constant K equal to $Mc$, the scalar field $\phi(x^\mu)$ satisfies the Klein-Gordon equation (32) and the mass is seen to arise as an effect of eliminating the internal variables. The negative sign of this term in Eq. (33) has no problem since $\phi(x^a)$ is not an observed physical field.

Single particle Dirac equation (13) represents a point particle with mass $m$; let us see if the mass term could be obtained from the internal motion in our model. First it has to be clarified that a spinor is a 2-component quantity that transforms as the irreducible representation $(\frac{1}{2}, 0)$ of the homogeneous Lorentz group, and a conjugate spinor transforms as $(0, \frac{1}{2})$ representation. Extending proper homogeneous Lorentz group to include space inversion the 2-component spinor representation is inadequate, and four components become necessary. Mathematical form of spinors was discovered by Elie Cartan in 1913 \cite{29}, and Pauli was the first to use them in physics \cite{31}. Dirac 4-spinor for electron theory in one of the representations of Dirac gamma matrices splits into a pair of 2-spinors, see \cite{14} for a nice historical background and physical insights and \cite{32} for a modern treatment.

In the absence of mass term the proposed equation for $\Psi(x^\mu, x^a)$ is
\begin{equation}
 (i\hbar \gamma^\mu \partial_\mu + if \gamma^a \partial_a ) \Psi =0
\end{equation}
We use the chiral or Weyl representation for gamma matrices given in the following $2 \times 2$ block diagonal form
\begin{equation}
\gamma^0 = \left(\begin{array}{cc} 0 & I \\  I & 0\\
                     
                    \end{array}\right)
\end{equation}
\begin{equation}
{\bf \gamma} =\left(\begin{array}{cc} 0 & \bf{\sigma} \\ -\bf{\sigma} & 0\\
                     
                    \end{array}\right)
\end{equation}
Suppose $\Psi$ comprises of a pair of 2-component spinors such that these are dependent on only one set of variables , i. e. $\Psi_1(x^\mu)$ and $\Psi_2 (x^a)$ then Eq. (34) splits into two equations
\begin{equation}
i\hbar (\partial_0 - {\bf \sigma.\nabla}) \Psi_1 = 0
\end{equation}
\begin{equation}
i f (\partial_{0(i)} + {\bf \sigma.\nabla}_{(i)}) \Psi_2 = 0 
\end{equation}
We have inserted $(i)$ in the operator index to signify internal variables in Eq. (38). This is, of course, a trivial separation. An interesting possibility exists to introduce oscillatory period corresponding to electron charge radius assuming factorization of $e^{i \omega t_i}$ in the 4-spinor $\Psi$. If the two 2-spinors are assumed as functions of $x^\mu$ and $x^a$ only then Eq. (34) gives rise to the following  2-spinor equations 
\begin{equation}
i\hbar (\partial_0 - {\bf \sigma.\nabla}) \Psi_1  - \frac{f\omega}{c} \Psi_1 = 0
\end{equation}
\begin{equation}
i f (\partial_{0(i)} + {\bf \sigma.\nabla}_{(i)}) \Psi_2 - \frac{f\omega}{c} \Psi_2  = 0 
\end{equation}
Using (18) and $\omega = 2 \pi m c^3 /e^2$ for the oscillation frequency for charge radius we obtain the mass-like term $mc$ in Eq. (39): thus the mass of free particle is related with internal oscillation.

Eq. (34) shows that $\Psi$ satisfies the second order wave equation (in both variables); similarly Eqs. (37) and (38) imply that $\Psi_1$ and $\Psi_2$ also obey the second order wave equation. However Eqs. (39) and (40) do not lead to second order Klein-Gordon equation for the spinors; it is, of course, expected because of the nonrelativistic approximation we made for separation. The last term on the left side of these equations is, therefore termed mass-like since it is easy to verify that plane wave solution in the rest frame gives rise to this interpretation.

Let us compare our results with Dirac theory. In Dirac equation (14) for the chiral representation of gamma matrices (35) and (36) the mass term mixes the two 2-spinors unlike the separation embodied in Eqs. (39) and (40) here. To get physical interpretation of four components of $\psi$ in Dirac equation a simplified assumption of plane wave solution and rest frame is often made \cite{14, 32} : two positive energy and two negative energy spinors are obtained corresponding to spin up and spin down states of electron and positron respectively where effectively only 2-component spinors are present. Phenomenologically in the present approach electron is represented by a 2-spinor for spin and one of the eigenstates of another for electric charge (interpreted as rotation in internal space). In general the assumed separation of $\Psi$ into $\Psi_1$ and $\Psi_2$ would not hold, and for realistic calculations an algorithm is necessary to relate the sub-quantum variables with some concrete variables. There could be a possibility of treating them as gauge transformations or these may be like  body-fixed and space-fixed ones \cite{5}. Further, the present work is limited to free particle case; in the next section we comment on aspects of interaction with electromagnetic field.

\section{\bf Discussion and Conclusion}

We must ensure that the present model of electron is not in conflict with the established physics. First note that the static Coulomb field finds re-interpretation, at the same time the problem of infinity that plagues point field theory is absent. The asymmetry between the sources for electric and magnetic fields is not of fundamental nature as electric charge itself is a flux quantum: it is only due to the small value of this flux quantum $e$ as compared to $hc/e$, and the rotatory motion of the vortex O that for large distances the observed field is electric field. Assumption of the rotating flux $e$ as a point charge discarding the vortex C and the associated spin and flux quantum $hc/e$ we get the classical picture in which charges and currents are the sources of the electromagnetic fields. In the standard classical electrodynamics spin of the electron does not play any role either in the description of the current flow or the Lorentz force law; the magnetic moment arises as a secondary effect. Obviously none of the experimental facts would be violated in our approach. However rather than seeking magnetic monopole, the elusive object not observed till date, here we have electron as a composite particle consisting of electric charge-like flux quantum and magnetic monopole-like flux quantum. Further the two electrons with opposite spin have distinct internal structure, therefore the spin polarization of current carrying electrons should show up in new electromagnetic phenomena. 

Is there something radically new in our work? To address this question we briefly review past attempts that sought alternatives so that our ideas are placed in proper perspective. Though as early as 1917 the role of magnetic energy in the spinning electron model of Abraham was discussed \cite{2}, in general magnetic field has been of lesser importance. Barut \cite{6} noted that,'It would have been strange if Nature provided magnetic forces just to be tiny corrections to the building principle of atoms ....'. In Barut's model the basic constituents of matter are assumed to be the stable particles: proton, electron, neutrinos and photon; and the only binding force is that of electromagnetic origin. It is shown that magnetic forces between the stable particles become very strong at short distances; the strong interaction between hadrons is interpreted as a dynamical spin-spin and spin-orbit force. Lepton-hadron distinction is not of significance, and muon is visualized as a magnetic excitation of electron due to the interaction of the anomalous magnetic moment with its own field.

Schwinger in 1969 speculated on a magnetic model of matter \cite{7}. A new ingredient in his model is the modification of Maxwell equations incorporating Dirac's monopole, however postulating a new species of particle: dyon. Dyon is a dual charged particle possessing electric charge with coupling constant $\alpha$ and magnetic charge with coupling constant $4/\alpha$. A tentative theory of hadrons is outlined noting that the force between magnetic charges is superstrong in comparison with the strong nuclear force. Leptons are not composite though it is suggested that neutrinos could belong to both lepton and hadron families.

Jehle in a remarkable series of papers in 1970s not only highlighted certain fundamental questions in the historical perspective but also formulated a new approach to electromagnetism and elementary particle physics \cite{8}. The standard electromagnetism is built on electric charge and its dynamics. Jehle puts forward the hypothesis that quantized flux $hc/e$ is fundamental and the electricity and electric properties are derived from it. A closed flux loop is an elementary object from which a manifold of loopforms is constructed. Electron and muon are represented by a single loop. Topology of linked and knotted flux loops is used to interpret quarks and classification of elementary particles.

A more radical though tentaive idea is that of quantum cohomology due to Post \cite{9}. In his book Post makes two main contributions: a strong criticism of the orthodox Copenhagen interpretation of quantum mechanics, and an alternative topological approach for fundamental physics. Unfortunately excessive and repetitive emphasis on the first has obscured the novelty of the topological approach. Electromagnetism as metric-free theory, the recognition of flux quantum as de Rham period integral, and the significance of topological torsion in 4-dimensions comprise Post's quantum cohomology. To avoid likely confusion with the term quantum cohomology, it has to be emphasized that Post's idea is entirely different than quantum cohomology of superstring literature \cite{33}. Electric charge is fundamental in Post's theory, and a 3-dimensional period integral for spin angular momentum proposed by Kiehn \cite{34} is a new addition to the well known 1-dimensional Aharonov-Bohm flux integral and 2-dimensional Ampere-Gauss charge integral. Electron and muon are represented by a trefoil knot and a 'preliminary cohomological classification' for electron, muon, neutrinos, pions and photon is presented.

The question arises as to why these attempts did not succeed. Schwinger in his paper highlights the speculative character of his ideas and at one place remarks that, 'However wide of the truth this hypothesis may be, it can serve to bring into better focus the nature of the quest for order and understanding that underlies the activity of the high-energy physicists'. It is possible that the ideas of Barut, Schwinger, Jehle and Post do not belong to the realm of the laws of Nature and have physical realization. This sort of conclusion would be rather premature since the mainstream physicists have not explored these ideas as vigorously as the most successful standard theories have been. Nevertheless let us have a critical look if there are weaknesses in these endeavours. One drawback common to them except that of the Post's work is that the new ideas were applied to the particle physics retaining the standard paradigm: the classification scheme based on the so called internal quantum numbers, conceptual framework of quantum field theory and quark models. Post argues for an alternative in which quark is not a legitimate object of physical reality; this, of course, would require tremendous effort to recast enormous knowledge in high energy physics in the alternative paradigm. Barut and Schwinger do not probe further if magnetism and magnetic charge could have deeper meaning than that following the Maxwell-Lorentz theory of electromagnetism. Jehle does take a step forward replacing the electric charge believed to be fundamental by elementary flux loop as a fundamental entity and also dispensing with magnetic monopole. I think there are two drawbacks in Jehle's approach that probably limited its scope. The derivation of electric field from quantized flux loop involves somewhat artificial introduction of a fraction of the quantized flux, denoted by F in his papers, since the loopform is assumed to spin at an angular velocity of $2mc^2/\hbar$ that corresponds to the Compton wavelength. The role of $\lambda_c$ is supported by the analysis of zitterbewegung given by Huang \cite{15}. The fraction F is linked with the quantized flux through
\begin{equation}
 F = \frac{e^2}{2 m c^2} \frac{1}{r}
\end{equation}
To obtain the electrostatic potential from the loop of quantized flux F has to be used. One may notice that F contains electron charge radius.
Though quantized charge is explained due to the quantized flux without postulating magnetic monopole, there is no explanation as to what charge is. Secondly Jehle drifts away to construct quark models as building blocks of matter.

Regarding quantum cohomology of Post it remains completely unexplored and ignored too. In my work \cite{22} a different idea than that of Post for topological torsion has been proposed for a new model of photon. Post treats flux as more fundamental than the magnetic moment, however electric charge is assumed fundamental elementary unit provided by Nature and independent of spacetime. Has this rigid assumption on electric charge blocked deeper insights from quantum cohomology?

Now it becomes straightforward to state the new elements in our approach: in contrast to a single flux quantum $hc/e$ in Jehle's model we have two flux quanta such that electric charge itself is a quantum of flux, the classical concept of charge in conjunction with flux $hc/e$ could be used to derive the notion of magnetic moment but it is not fundamental, and finally the magnetic flux itself is a derived concept from the vorticity or the circulation of the spacetime fluid vortex. The concept of electric charge proposed here is radically different than that of Post since spacetime rotation manifests as charge while according to Post charge is independent of spacetime. Electron is a composite of two vortices or two flux quanta - it is akin to Schwinger's dyon. The concept of composite particles discussed in the literature on fractional quantum Hall effect should not be confused with our electron model. It has to be emphasized that in these theories the flux quantum of a vortex is created by the application of the external magnetic field on a 2-dimensional electron system; for further discussion and application to spintronics see \cite{35}.

Fluid or hydrodynamical approach to electromagnetism in the nineteenth century, and to quantum theory in the 1920s is
well known. The present ideas in which electric charge is given a mechanical interpretation could stimulate revival of the fluid dynamical
 paradigm for fundamental physics: in \cite{36} the electromagnetic field tensor has been interpreted as angular momentum tensor
of photon fluid, the representation of sources in terms of flux integrals would render this description to a completion. Instead of a point
 charge what we have is a flux integral for electric charge, therefore the divergence problem will not arise. Quantized vortices
are best treated as topological objects making geometric and topological rendition of the electromagnetic phenomena quite natural.

The concept of vortex discussed here has to be made mathematically precise that we propose to do in near future, however a plausible approach is outlined as follows. A finite-sized wavepacket rather than plane wave solution is needed to interpret spin as some kind of rotation. Huang constructs \cite{15} a Gaussian wavepacket to discuss the spatial extension in the solution of Dirac equation and obtains spin from zitterbewegung. He rightly emphasizes that electron motion within a tube of the diameter of the order of $\lambda_c$ cannot be determined and only an approximate rough picture could be formed. Since each of the four components of Dirac spinor in Dirac equation also satisfies the second order Klein-Gordon equation, considering one of the components it immediately follows that a vortex solution could be obtained for it: line singularities in the solution of the scalar wave equation called wave dislocations \cite{37} for which the amplitude is zero and the phase is indeterminate. For example, at time $t=0$, the gradient of the phase of scalar wave dislocation \cite{36} in cylindrical coordinate system $(\rho, \phi, z)$ is
\begin{equation}
{\bf \nabla} (kz + \ell \phi + \Phi) = k\hat{z} + \frac{\ell}{\rho} \hat{\phi} + {\bf \nabla} \Phi
\end{equation}
One may interpret $\ell$ as topological charge of phase singularity; Brambilla et al \cite{38} present a nice hydrodynamical analogy such that the phase is interpreted as the velocity potential and the dislocation as a vortex. Application of these ideas in optical beams is well known for vector wave equations \cite{36}; it would be interesting to investigate them for Dirac equation, e. g. replacing Huang's Gaussian solution by the Laguerre-Gaussian wavepackets. While interference between negative and positive energy solutions in a region less than $\lambda_c$ would give the usual zitterbewegung in Dirac theory, the 2-spinor Eqs  (37) and (38) would have no such problem. 

We conclude the paper with following remarks. 1) The proposition that electric charge is a quantum of flux, and the conceptual framework for sub-quantum theory applicable to the structure of the electron in a region of the order of its Compton wavelength constitute the principal contribution of the present work. In a simplified Dirac picture the sub-quantum conditions using Eq. (26) are shown to lead to the anomalous magnetic moment of electron. A universal constant $f$ defines the sub-quantum rules just as $\hbar$ determines quantum conditions.

The past efforts to construct fractional spin of electron \cite{39} and reconcile the flux quantization in the units of $hc/e$ \cite{40} in the light of our electron model had a very limited success. Now the sub-quantum theory in the internal space allows spinor representation for spin $f/2$ that is numerically a fraction of $\hbar /2$. The quantum of flux following the usual Aharonov-Bohm phase prescription in sub-quantum theory can be obtained to be $e$. The integral relation $\frac{e}{fc} \oint {\bf A.dl} = 2\pi$ gives rise to the quantum of flux $\int {\bf B.dS} =e$.
This makes it  possible to envisage a 2-vortex structure of electron as an abstraction of flux quanta.

2) Discussion on discrete symmetries and construction of vortex solutions need further insights and mathematical advancement. In the Dirac framework itself there are interesting open questions, see e. g. \cite{5, 12, 13}. Could there be a geometrical model that incorporates both Dirac spinor aspect and a picture of spinning extended object? An interesting idea proposed by Burinskii \cite{41} could prove useful in this connection. Two of the important features in this work \cite{41} seem attractive: first the electron has a natural extended spacetime structure, and second there exists a correspondence between spinor-twistor in Kerr geometry and spinor in Dirac theory. Recalling the geometrical interpretation of a spinor a la Cartan and a tentative electron model in terms of Kerr geometry in 2+1 dimension discussed on p. 198 in \cite{4} the new development reported in \cite{41} deserves serious consideration.

3) A thorough approach to describe interaction with external fields has not been developed though a simple equation to discuss interaction in internal space has been given vide Eq. (26). A notable point in our theory is that we have an additional sub-quantum picture, therefore it would be of interest to examine two limits in analogy with the conventional treatment, namely the quantum limit of sub-quantum physics and semiclassical limit of the resulting quantum theory. We propose to address this issue in near future.

4) Though we have not given a rigorous treatment of internal variables it would be interesting to examine renormalization in QED afresh. Quantum vacuum in QED invokes electron, positron creation and a kind of electron-positron cloud within a region of the order of $\lambda_c$. In one of the methods counter terms are added to the QED Lagrangian and relationships are introduced between the bare and the observed quantities \cite{14, 32} to obtain ultimately finite results. In our model there does not exist electron-positron cloud, and the electron has extension in space such that of the two length scales electron charge radius may be viewed roaming inside $\lambda_c$; note the relation (30). Let us consider Eqs. (15) and (26) and assume single 4-spinor $\Psi$ as a function of both sets of variables then a simple action function becomes
\begin{equation}
 S = \int [\bar{\Psi} \gamma^\mu (i\hbar \partial_\mu - \frac{e}{c} A_\mu) \Psi  + \bar{\Psi} \gamma^a (if\partial_a - \frac{e_s}{c} A_a) \Psi] d^4x d^4x_{(i)}
\end{equation}
Note that we have used charge parameter $e_s$ for internal space. Could the second term in the integrand of (43) play the role of counter terms in QED? If this guess is proved correct it surely would help us resolve the most outstanding problem in QED: the foundation of the renormalization method.

Acknowledgments

I am grateful to one of the Referees for a painstaking review, useful suggestions and reference \cite{41}. The library facility of Banaras Hindu University is acknowledged.

\end{document}